\documentclass[a4paper]{jpconf}
\usepackage{graphicx}
\usepackage{multicol}
\usepackage{subfigure}
\usepackage{amsmath}
\begin{document}
\title{Implementation of an upward-going muon trigger for indirect dark matter searches at the NO$\nu$A far detector}

\author{R. Mina$^1$, M. J. Frank$^1$, E. Fries$^1$, R. C. Group$^{1,2}$, A. Norman$^2$, and I. Oksuzian$^1$}

\address{$^1$University of Virginia, Charlottesville VA, USA}
\address{$^2$Fermilab National Accelerator Laboratory, Batavia IL, USA}

\ead{ram2aq@virginia.edu}


\begin{abstract}
The NO$\nu$A collaboration has constructed a 14,000 ton, fine-grained, low-Z, total absorption tracking calorimeter at an off-axis angle to an upgraded NuMI neutrino beam. This detector, with its excellent granularity and energy resolution and relatively low-energy neutrino thresholds, was designed to observe electron neutrino appearance in a muon neutrino beam, but it also has unique capabilities suitable for more exotic efforts. In fact, if an efficient upward-going muon trigger with sufficient cosmic ray background rejection can be demonstrated, NO$\nu$A will be capable of a competitive indirect dark matter search for low-mass WIMPs. The cosmic ray muon rate at the NO$\nu$A far detector is about 100 kHz and provides the primary challenge for triggering and optimizing such a search analysis. The status of the NO$\nu$A upward-going muon trigger is presented.
\end{abstract}

\section{Introduction}
WIMPs captured by the gravitational field of the Sun that are slowed
through collisions with solar matter can accumulate in the solar core.
There, WIMP annihilation may produce neutrinos with much larger energy
than solar neutrinos.  The signal would be an excess of high-energy
($>0.5$\,GeV) neutrino events pointing back to the Sun~\cite{Hagelin:1986gv,Buckley:2013bha}.  The cleanest
signature at NO$\nu$A will be from $\nu _{\mu}$ CC events producing
upward-going muons that can be reconstructed in the NO$\nu$A detector.
The large and unique NO$\nu$A far detector, with its excellent granularity
and energy resolution, and relatively low-energy neutrino thresholds,
is an ideal tool for these indirect dark matter searches.

At NO$\nu$A, the neutrino analyses simply store events synchronous
with the NuMI beam.  For non-beam exotic physics searches, so-called
data-driven triggers~\cite{Fischler:2012zz} are required to select events of interest. 

Only the upward-going flux will be considered in order to suppress
the cosmic-ray background.  The downward-going muon rate in the NO$\nu$A
far detector is approximately 100,000 Hz.  We expect to keep the
upward-going muon trigger rate to about 10 Hz or less, so a rejection
of at least four orders of magnitude is required by the trigger. Of course, this rejection must be accomplished while keeping
the acceptance for upward-going muons relatively high. 

 The neutrino flux from dark matter annihilation is model
 dependent; however, energies from $\sim$0.5\,GeV to many TeV should
 be detected with high acceptance.  For high-mass signal hypothesis,
 NO$\nu$A will not be able to compete with the high acceptance of the IceCube
 detector~\cite{Aartsen:2012kia}.  For lower-mass scenarios (below
 $\sim$20 GeV) the Super-Kamiokande experiment currently has the
 best sensitivity~\cite{Tanaka:2011uf,Choi:2015ara}.  If an efficient upward-going
 muon trigger and sufficient cosmic ray background rejection can be
 achieved, NO$\nu$A will be competitive with Super--Kamiokande
 for WIMP mass hypotheses below 20 GeV/c$^2$.

One advantage that NO$\nu$A has compared to past experiments that
performed similar searches for dark matter annihilation is the
relatively low energy threshold for muons.  A 1 GeV muon track travels
approximately 5 meters in the NO$\nu$A detector resulting in an energy threshold
well below 1 GeV.  The challenge for the dark matter search is
triggering efficiently on these low-energy muons.  For shorter track
lengths, the timing information will not be as powerful for rejecting
downward-going backgrounds.  Using stopping or fully-contained events
and using the top and sides of the detector to veto downward-going
events can provide an additional two orders of magnitude
rejection. 

In this note we focus on using the timing information from all of the hits on a track to reject the downward-going muon background and efficiently select upward-going events.

\section{Hit time estimates}

A trigger for upward-going muons based on timing information
required a minor upgrade to the readout of the NO$\nu$A far detector.
This upgrade to the so-called ``multipoint'' readout occurred on
September 11, 2014, and resulted immediately in a single-hit timing
resolution of about 25 ns (note that the timing resolution with the previous
algorithm was about 125 ns, so this is a significant improvement).
With dozens of hits per track, it is possible to reject
downward-going muons by many orders of magnitude using hit timing information alone.

To resolve the directionality of the muon track, the upward-going muon
trigger takes advantage of the timing information from each individual
hit in the reconstructed tracks. The tracks are reconstructed using the
Hough transform algorithm, and are required to match in both XZ and YZ
views. We start from the hit with lowest $y$ cell value, $y_0$,
in the track in the YZ view. The measured time of the corresponding hit is
defined as $T_0$. The observed and expected time of each hit on the
track in the YZ view is therefore:

\begin{equation}
  \label{eq:Ytimes}
  \begin{split}
  T_{obs} = TDC_{y_i} \cdot 15.625 - T_0 \\
  T_{exp} = TOF_\mu \frac{y_i - y_{0}}{y_{1} - y_{0}}
  \end{split}
\end{equation}

Similarly, for the XZ view:

\begin{equation}
  \label{eq:Xtimes}
  \begin{split}
  T_{obs} = TDC_{x_i} \cdot 15.625 - T_0 \\
  T_{exp} = TOF_\mu \frac{x_i - x_{0}}{x_{1} - x_{0}},
  \end{split}
\end{equation}

 where $x_i$ and $y_i$ are the cell numbers in XZ and YZ view, and
$TDC_{x(y)_i}$ is the time measurement in TDC units, which is converted to ns
using the factor of 15.625 ns/TDC.
$TOF_{\mu}$ is the time-of-flight of the muon track defined as:

\begin{equation}
  TOF_\mu = L/29.97,
\end{equation}
 where $L$ is track length in cm, and 29.97cm/ns  is the expected speed
assuming that the muon is relativistic. 

Since we require that each track is reconstructed and matched in both
views, ($x_0$; $y_0$) and ($x_1$; $y_1$) must correspond to the lowest
and highest points of the track respectively. In addition, we can
estimate the missing coordinate for a particular hit in either view using 3D requirement. For the YZ view,
track coordinates can be calculated as such:

\begin{equation}
  \begin{split}
  x = \frac{x_1 - x_{0}}{z_{1} - z_{0}} \cdot (z - z_0) + x_0 =
  \frac{x_1 - x_{0}}{y_{1} - y_{0}} \cdot (y - y_0) + x_0 \\
  y = c_0 + c_w \cdot c \\
  z = p_0 + p_w \cdot p \\
  \end{split}
\end{equation}

Similarly, for the XZ view:

\begin{equation}
  \begin{split}
  x = c_0 + c_w \cdot c \\
  y = \frac{y_1 - y_{0}}{z_{1} - z_{0}} \cdot (z - z_0) + y_0 =
  \frac{y_1 - y_{0}}{x_{1} - x_{0}} \cdot (x - x_0) + y_0 \\
  z = p_0 + p_w \cdot p, \\
  \end{split}
\end{equation}

where $c_w = 3.97$ cm and $p_w = 6.65$ cm are the widths of detector cells and planes. The cell and plane with id=0 have coordinates $c_0 =
-759.50$ cm and $p_0 = 4.57$ cm.

Since for each hit in each view we can estimate (x; y; z) coordinates,
we can calculate the distance from the hit to the APD readout end. The further
the hit is located from the readout the longer it takes for the light
to propagate and be detected by the APD. We are interested in the hit
time of the muon passing through the extrusion, so we have to
correct for the light propagation time in the fiber. The speed of light in the
fiber is measured to be 15.3 cm/ns.

\subsection{Multipoint fine timing}

The light level in each channel in the NO$\nu$A detector is independently sampled every 500 ns.
The electronic response to an incident particle depositing energy in a cell
can be parameterized in terms of two intrinsic timing values ($T_F$ and $T_R$), the number of photoelectrons ($p$),
and a timing ``offset'' ($t_0$), or the elapsed time between a read-out and the time of incidence of the particle:

\begin{equation}
		f(t) = \alpha pe^{-(t-t_0)/T_F}(1-e^{-(t-t_0)/T_R})
\end{equation}

Here, $\alpha$ is a proportionality factor that does not affect the timing fit. The parameters $T_F$ and $T_R$ correspond to the intrinsic falling
and rising time of the response curve, respectively. As such, they are 
approximately known. For the purpose of determining hit timing, the
parameter of note is $t_0$. By performing a simple $\chi^2$ minimization,
the data-preferred value of $t_0$ can be extracted from multiple readouts
on a single channel. For the purposes of the trigger, where hit processing time
must be minimized, fit results were pre-calculated and tabulated such that
the computationally expensive minimization need not be repeated
for each individual hit.

\begin{figure}[hbt]
\centering
\includegraphics[width=2.0in]{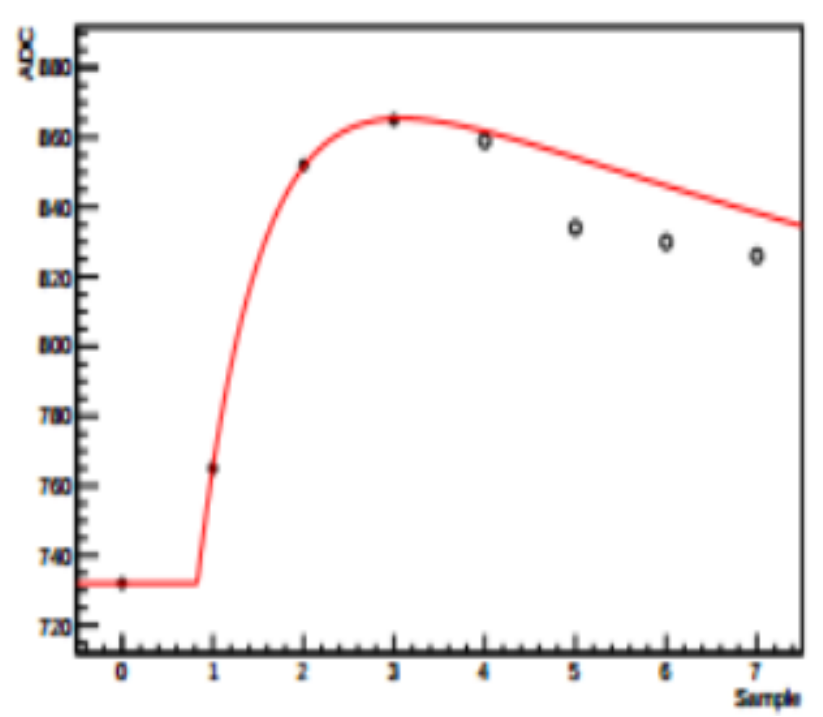}
\caption{\label{fig:TimingFit} An example of fitting the electronics 
    response curve to multiple readouts from a single cell hit. The 
    time coordinate of the inflection point where the curve
	begins to rise is the fitted parameter $t_0$.}
\end{figure}
 
\begin{figure}[hbt]
 \centering
 \includegraphics[width=6.0in]{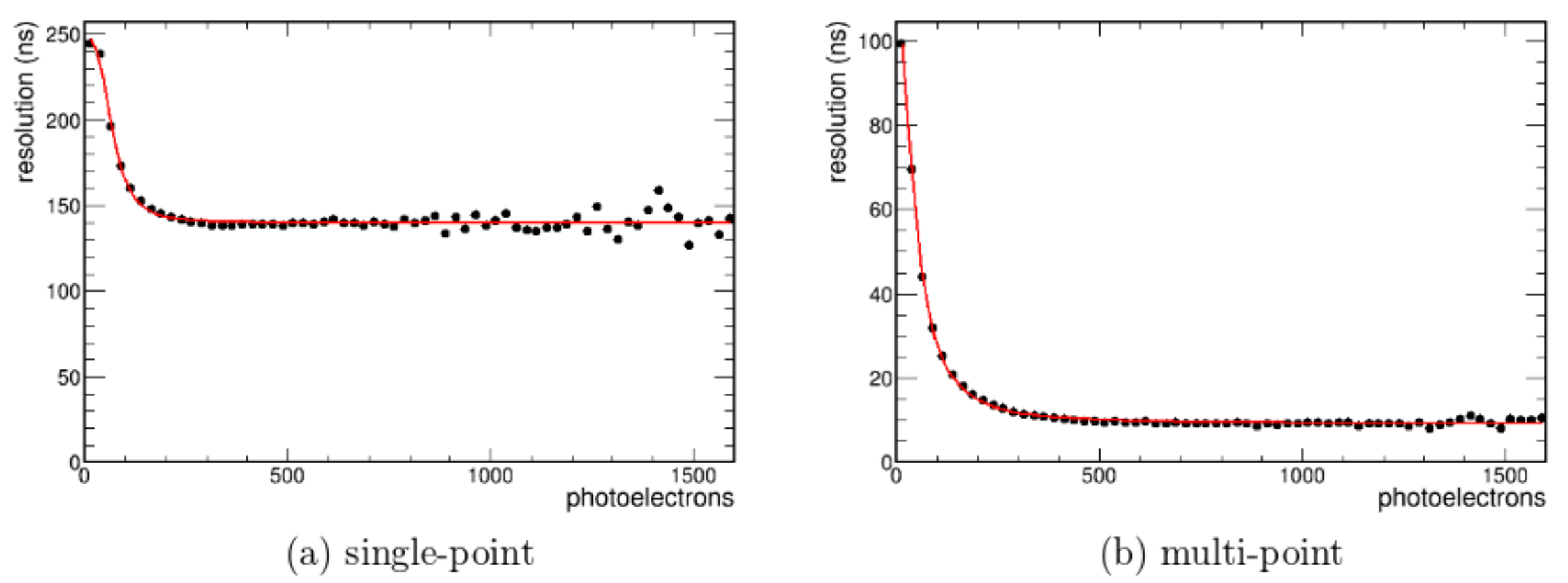}
 \caption{\label{fig:DeltaT}  Single-hit timing resolution as
   observed in NO$\nu$A far detector data with four-point readout, 
   before (left) and after (right) fine timing implementation.
   See Ref.~\cite{evan} for more details.}
\end{figure}

Each time measurement has an uncertainty, which varies with the amount 
of energy deposited. The time uncertainty on a given hit from a reconstructed
muon track affects the determination of track directionality, so a 
parameterization of uncertainty in terms of energy deposition is 
necessary for the timing-based trigger. Single-hit time resolution is 
plotted against energy 
deposition in Fig.~\ref{fig:DeltaT}. For high energy hits the $\Delta{t}$ 
is measured to be  approximately 10 ns in the data using the four-point 
readout scheme, which is consistent with that observed in simulation 
~\cite{evan}.

\section{Log likelihood ratio}

We can use equations~\ref{eq:Ytimes} and~\ref{eq:Xtimes} to produce
the distribution of the expected v/s observed time for each
track.

\begin{figure}[hbt]
  \centering
  \includegraphics[width=4.0in]{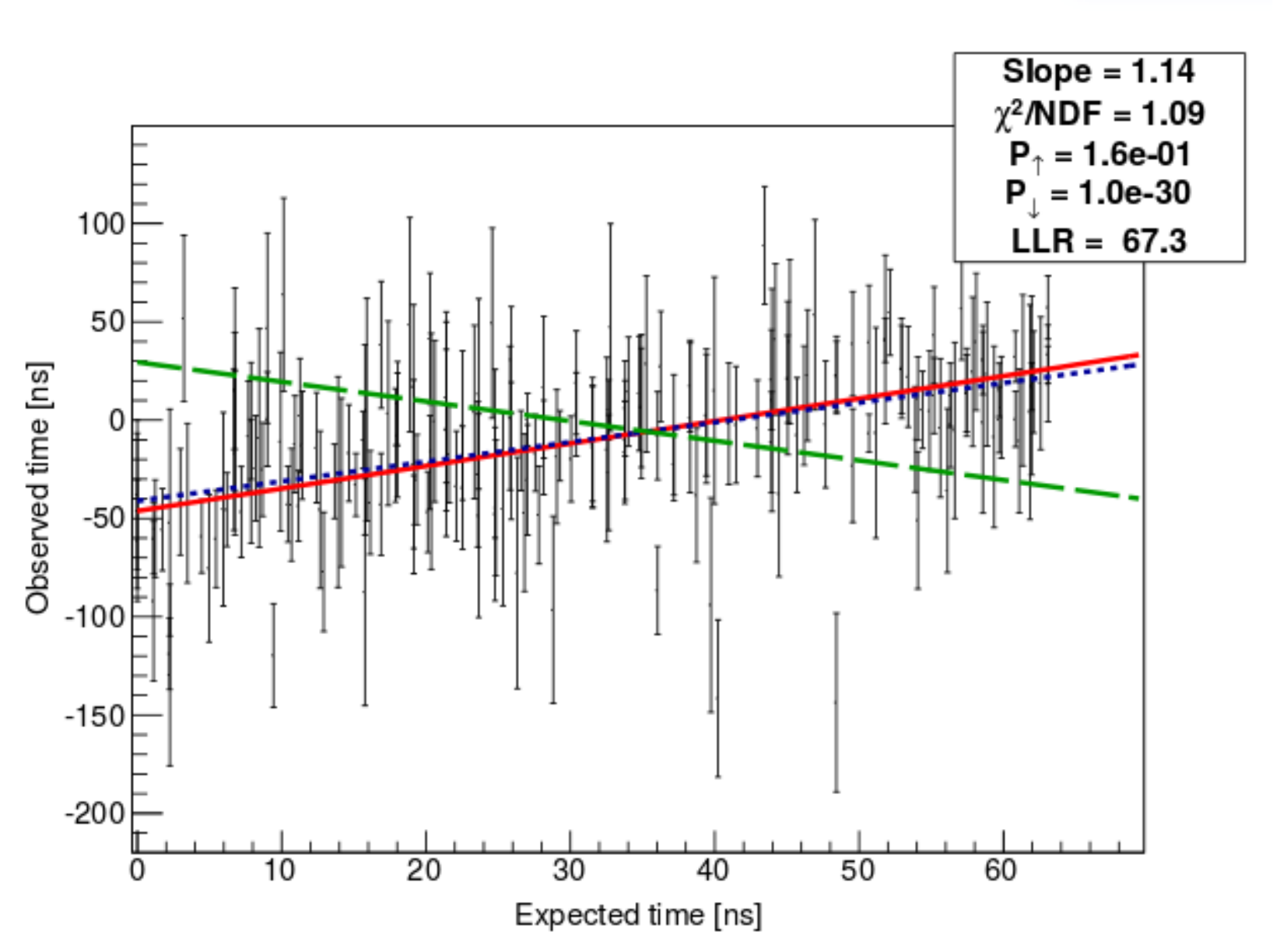}
  \caption{\label{fig:track_example}  The expected versus observed time
    distribution for an upward-going muon track
    reconstructed in the NO$\nu$A far detector, using fine timing. The linear unconstrained fit (red solid
    line) has slope value close to ``1''. The fit with the upward-going track
    hypothesis (slope = 1) is shown as the blue-dashed line.  The fit with the downward-going hypothesis is shown in the green dashed line and has a very poor probability. }
\end{figure}

\begin{figure}[hbt]
  \centering
  \includegraphics[width=6.0in]{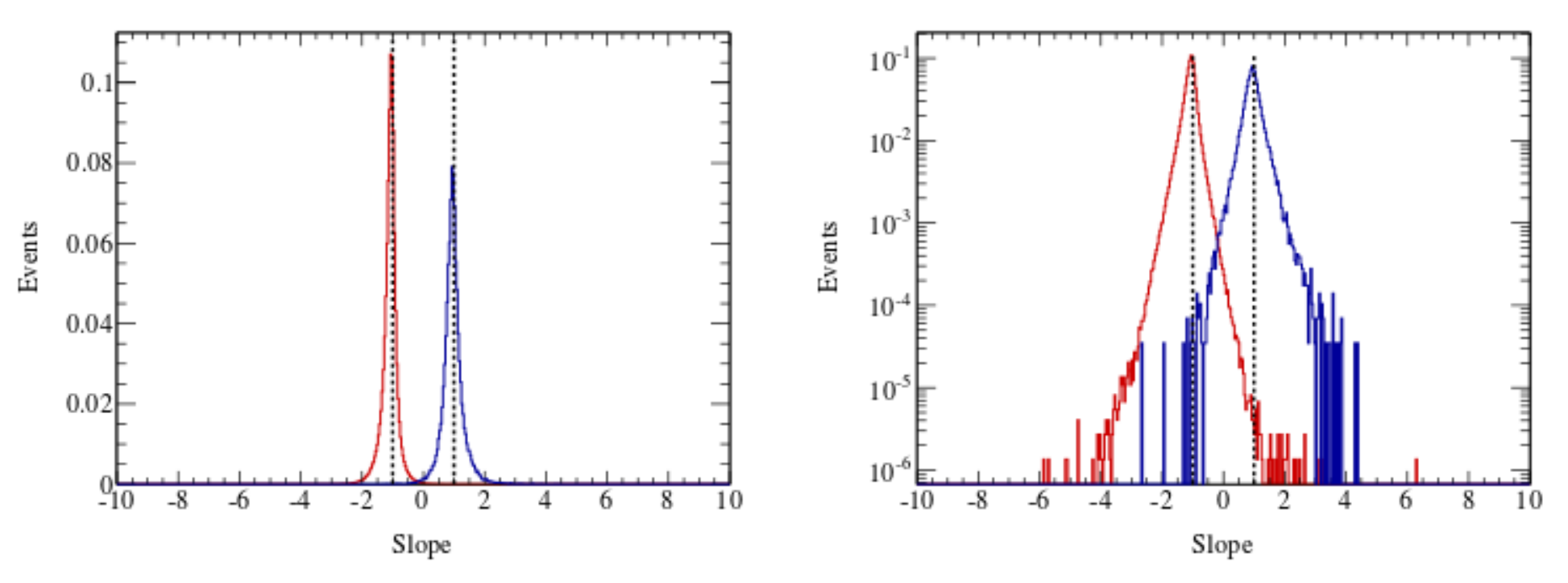}
  \caption{\label{fig:disc_slope}  The slope distributions
    for cosmics (red) and {\sc WimpSim} (blue) MC samples.}
\end{figure}

An example of expected v/s observed time distribution is shown in
Fig.~\ref{fig:track_example}. The distribution is produced using a
reconstructed upward-going muon track simulated with {\sc
  WimpSim}~\cite{Blennow:2007tw,wimpsim}. As can be seen, the points
follow a rising trend with a slope value consistent with the
upward-going track hypothesis. It is clear from the figure that the
fitted slope value can be used to estimate the muon direction (up or
down). As shown in Fig.~\ref{fig:disc_slope}, the slope values for
cosmics and {\sc WimpSim} MC samples are consistent with the downward- and
upward-going hypothesis, respectively.  In the relativistic limit, it
is safe to assume that there are only two options for the slope
values. Therefore, we can fit the time distribution on
Fig.~\ref{fig:track_example} with fixed values of slopes. For the
upward-going track the fit with the slope constrained to ``1'' results
in a good $\chi^2$ probability value of the fit,
$P_{\uparrow}$. However the fit with slope of ``-1'' yields a low
probability value, $P_{\downarrow}$ . Using the probability values
from the fits with the fixed slope value, we can form a log-likelihood
ratio (LLR):

\begin{equation}
 LLR = Log(\frac{P_{\uparrow}}{P_{\downarrow}})
\end{equation}
 
\begin{figure}[hbt]
  \centering
  \includegraphics[width=4.0in]{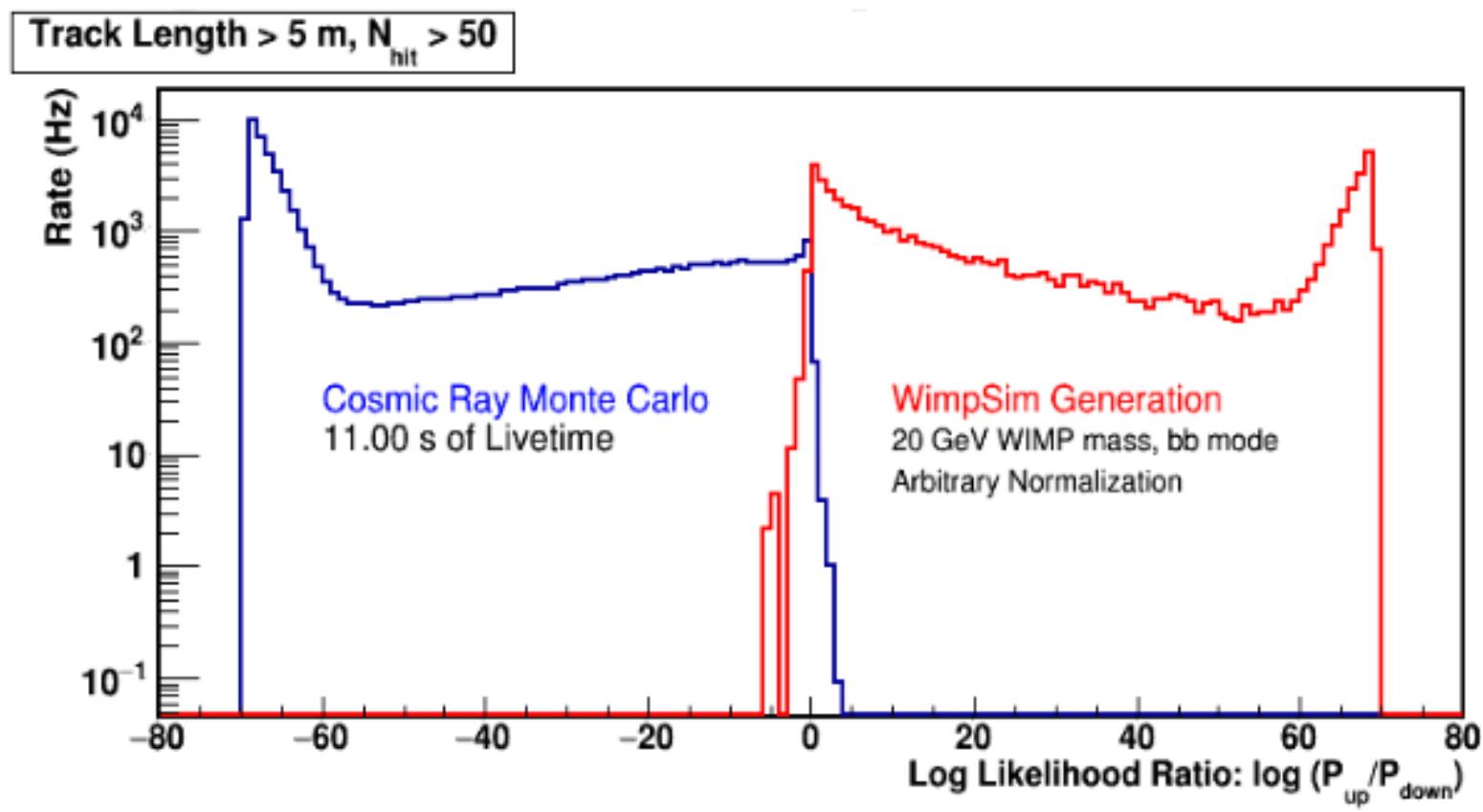}
  \caption{\label{fig:disc_llr}  The LLR distributions
    for cosmics (red) and {\sc WimpSim} (blue) MC samples. Note that only
    tracks longer then 5 m and with more than 50 hits are included.}
\end{figure}

The LLR distributions for the cosmic and {\sc WimpSim} MC samples are shown
in Fig.~\ref{fig:disc_llr}.  From this distribution, it is clear that a
cut on LLR slightly above zero will reduce the cosmic background by
the desired amount while preserving a high signal acceptance.  Note
that the {\sc WimpSim} sample used is for dark matter with a 20 GeV mass
annihilating through the $b\bar{b}$ channel.  As such the neutrinos from
the b-meson decay produce muons which, on average, have a much lower
energy compared to the cosmic ray muons.  This explains why the LLR
for the signal has a larger component close to zero than the cosmic
sample.

The LLR yields better performance for cosmic background rejection for the same
signal acceptance in the regime where the cosmic rejection is
sufficient (at least four orders of magnitude), compared to a cut on
the best-fit slope. For example, for a
signal acceptance of 0.7 the background rejection is about a factor of
three better for the LLR.  At this point the MC predicts background
rejection of close to five orders of magnitude.  In addition to being
a more powerful discriminator as observed in the MC studies, the LLR
estimator is more robust to mis-reconstructed tracks which will be an
important feature in real data.  Since mis-reconstructions will result
in time distributions that follow neither the upward- nor the
downward-going hypothesis, the result of mis-reconstruction
will yield LLR values close to ``0'', and not values consistent with a
high-probability for being upward-going.

\section{Conclusions}
A timing-based upward-going muon trigger was implemented for the NO$\nu$A far
detector and was deployed in November 2014. Triggering at $<10$Hz, the
algorithm suppresses cosmic ray muons by five orders of magnitude.

\begin{figure}[hbt]
	\centering
	\includegraphics[width=4in]{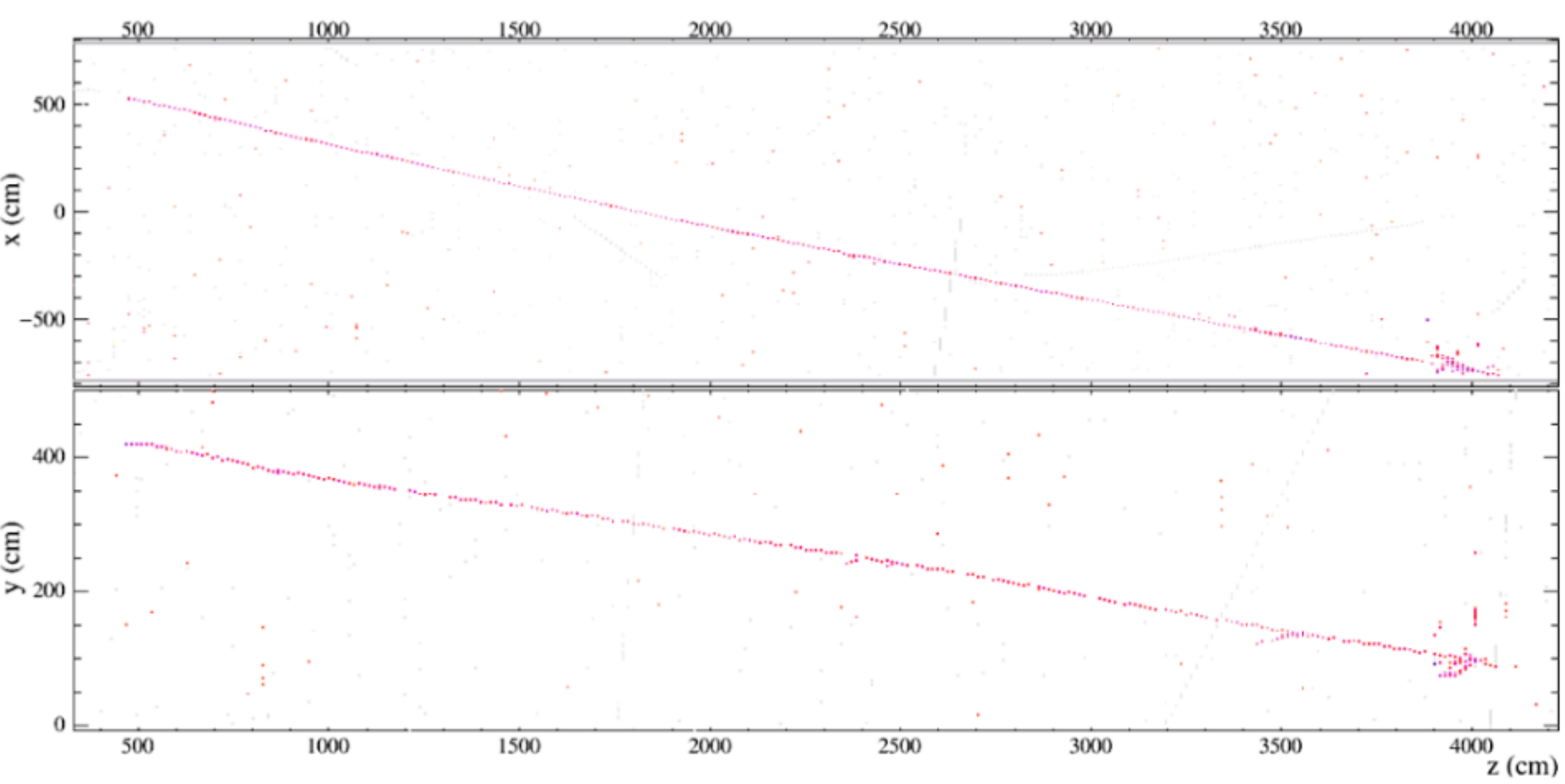}
	\includegraphics[width=2.0in]{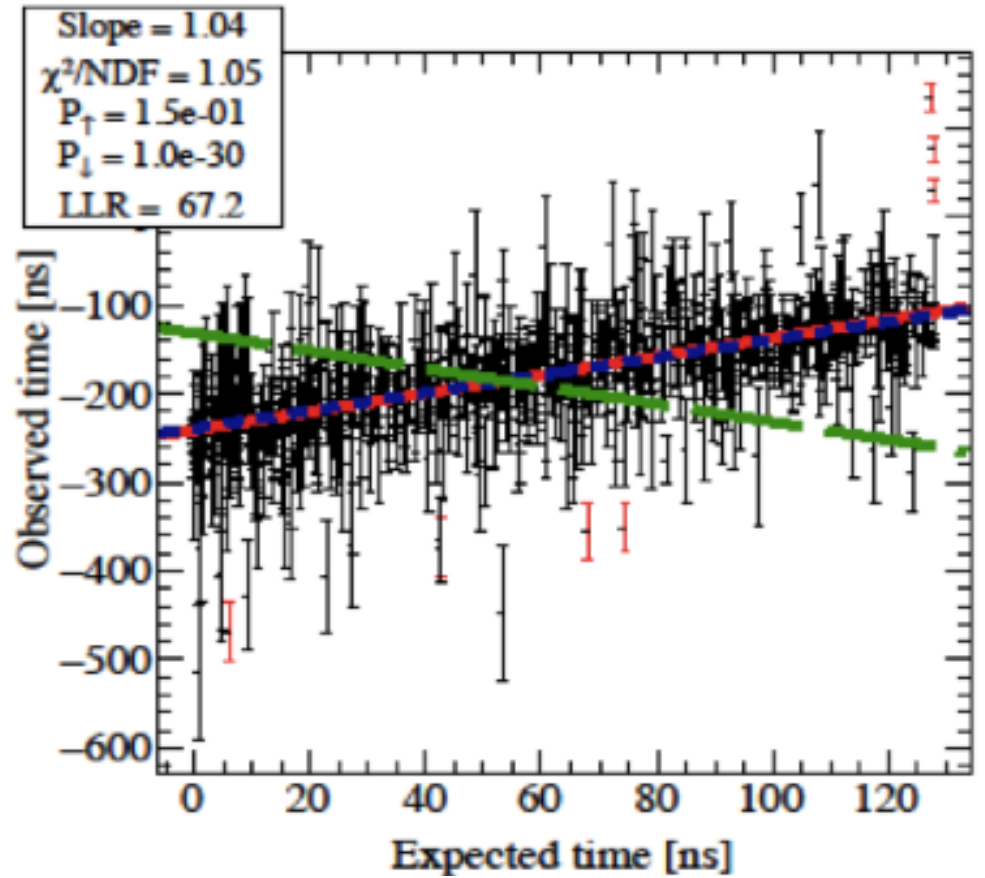}
	\caption{\label{fig:conc_example} On the left is a display of
	a triggered event that is a strong candidate, based on its
	topology, for an upward-going muon.  The activity at the
	bottom right indicates a CC scattering interaction. The
	curving at the other end probably indicates that the muon
	ranged out. There is also evidence for a Michel electron,
	based on the timing information on the right.  This event
	confirms that the LLR algorithm is successfully selecting
	upward-going muons in the data.}
\end{figure}

As can be seen in Fig.~\ref{fig:conc_example}, events with probable
Michel electrons and contained vertices have been used to confirm 
upward-going muons in the triggered sample.

Atmospheric neutrinos generated on the other side of the Earth are also
capable of producing upward-going muons in the detector. These events represent
an irreducible background in this search. The only method of discriminating
atmospheric neutrino events from WIMP events is to reconstruct the
directionality of the incident neutrinos, which has yet to be attempted in NO$\nu$A.  

The accumulating data sample opens the door to a program to study atmospheric neutrinos and sets the stage for a competitive dark matter search by the NO$\nu$A experiment.  

\ack
This conference presentation was made possible by a grant from the
University of Virginia College of Arts and Sciences. Additional
financial support was provided by the Jefferson Trust, the UVa Physics
Department, and the Fermilab Particle Physics Division. The authors
also acknowledge that support for this research was carried out by the
Fermilab scientific and technical staff. Fermilab is Operated by Fermi
Research Alliance, LLC under Contract No.~DE-AC02-07CH11359 with the
United States Department of Energy.  The University of Virginia
particle physics group is supported by DE-SC0007838.

\end{document}